\documentclass[12pt]{article}
\usepackage{graphicx}
\begin{document}

\begin{center}

{\Large \bf  Poincar\'e Symmetry from \\[1ex]
Heisenberg's Uncertainty Relations}

\vspace{16mm}

Sibel Ba{\c s}kal  \\

Department of Physics, Middle East Technical University, 06800 Ankara, Turkey \\[5ex]

 Young S. Kim \\
Center for Fundamental Physics, University of Maryland College Park,\\
Maryland, MD 20742, USA \\[5ex]

 Marilyn E. Noz \\
Department of Radiology, New York University, New York, NY 10016, USA

\end{center}

\vspace{10mm}

\abstract{It is noted that the single-variable Heisenberg commutation relation
 contains the symmetry of the $Sp(2)$ group which is isomorphic to the
 Lorentz group applicable to one time-like dimension and two space-like
 dimensions, known as the $SO(2,1)$ group.  According to Paul A. M. Dirac,
 from the uncertainty commutation relations for two variables, it possible
 to construct the de Sitter group $SO(3,2)$, namely the Lorentz group
 applicable to three space-like variables and two time-like variables.
 By contracting one of the time-like variables in $SO(3,2)$, it is possible,
 to construct the inhomogeneous Lorentz group $ISO(3,1)$ which serves as the
 fundamental symmetry group for quantum mechanics and quantum field theory
 in the Lorentz covariant world.  This $ISO(3,1)$ group is commonly known
 as the Poincar\'e group.  }

\newpage

\section{Introduction}
As early as in 1927~\cite{dir27}, Paul A, M. Dirac considered the problem
of extending Heisenberg's uncertainty relations to the
Lorentz-covariant world.
In 1945~\cite{dir45}, he attempted to construct the Lorentz
group using the Gaussian wave function.
In 1949~\cite{dir49},
Dirac pointed out the task of constructing relativistic dynamics
is to construct a representation of the inhomogeneous Lorentz group.
He then wrote down the ten generators of this group and their closed set
of commutation relations.   This set is known as the Lie algebra of
the Poincar\'e group.

In 1963~\cite{dir63}, Dirac considered two coupled harmonic oscillators
and constructed an algebra leading to the Lie algebra for the $SO(3,2)$
de Sitter group, which is the Lorentz group applicable to three space
dimensions and two time-like variables.

From the mathematical point of view, it is straight-forward to contract
one of those two time-like dimensions to construct $ISO(3,1)$ or the
Poincar\'e group.  This is what we present in this paper.  However,
from the physical point of view, we are deriving the Poincar\'e symmetry
for the Lorentz-covariant quantum world purely from the symmetries of
Heisenberg's uncertainty relations.

In Sec.~\ref{sp2}, it is noted that a one-dimensional uncertainty relation
contains the symmetry of the $Sp(2)$ group in the two-dimensional phase
space.  It is pointed out that this group, with three generators, is isomorphic
to the Lorentz group applicable to two space dimensions and one time variable.
We can next consider another set with three additional generators.

In Sec.~\ref{2osc}, we write those Heisenberg uncertainty relations in
term of step-up and step-down operators in the oscillator system.  It
is then possible to consider the two coupled oscillator system
with the ten generators constructed by Dirac in 1963~\cite{dir63}.
It is gratifying to note that this oscillator system can serve as the basic
language for the two-photon system of current interest~\cite{yuen76,yurke86}.

In Sec.~\ref{contraction}, we contract one of the time-like variables in
$SO(3,2)$ to arrive at the inhomogeneous Lorentz group $ISO(3,1)$ or the
Poincar\'e group. In Sec.~\ref{remarks}, we give some concluding
remarks.

\section{Sp(2) Symmetry for the Single-variable Uncertainty Relation}\label{sp2}
It is known that the symmetry of quantum mechanics and quantum field
theory is governed by the Poincar\'e group~\cite{dir49,dir62}.  The
Poincar\'e group means the inhomogeneous Lorentz group which includes
the Lorentz group applicable to the four-dimensional Minkowskian
space-time, plus space-time translations~\cite{wig39}.

The question is whether this Poincar\'e symmetry is derivable from
Heisenberg's uncertainty relation, which takes the familiar form
\begin{equation}\label{101}
   \left[x_{i}, p_{j}\right] = i \delta_{ij}.
\end{equation}
There are three commutation relations in this equation. Let us choose
one of them, and write it as
\begin{equation}\label{102}
   [x, p] = i .
\end{equation}
This commutation relation possesses the symmetry of the Poisson bracket in
classical mechanics~\cite{arnold78,stern84}.  The best way to address this property
is to use the Gaussian form for the Wigner function defined in the phase
space, which takes the
form~\cite{hkn88,kiwi90ajp,knp91}
\begin{equation}\label{103}
W(x,p) = \frac{1}{\pi} \exp\{-\left(x^2 + p^2\right)\}.
\end{equation}
This distribution is concentrated in the circular region around the origin.
Let us define the circle as
\begin{equation}\label{105}
x^{2} + p^{2} = 1.
\end {equation}
We can use the area of this circle in the phase space of $x$ and $p$ as the
minimum uncertainty.  This uncertainty is preserved
under rotations in the phase space:
\begin{equation}\label{107}
\pmatrix{\cos\theta & -\sin\theta \cr \sin\theta & \cos\theta }
\pmatrix{ x \cr p },
\end{equation}
as well as the squeeze of the form
\begin{equation}\label{108}
\pmatrix{e^{\eta} &  0 \cr 0 &  e^{-\eta} }
\pmatrix{x \cr p } .
\end{equation}

The rotation and the squeeze are generated by
\begin{equation}\label{109}
J_{2} = - i\left(x \frac{\partial}{\partial p} - p\frac{\partial}{\partial x } \right),
\qquad
K_{1} = -i \left(x\frac{\partial}{\partial x} - p \frac{\partial}{\partial p} \right),
\end{equation}
respectively.
If we take the commutation relation with these two operators, the result is
\begin{equation}\label{111}
\left[J_{2}, K_{1}\right] = -i K_{3},
\end{equation}
with
\begin{equation}\label{113}
K_{3} = -i \left(x\frac{\partial}{\partial p} +
p \frac{\partial}{\partial x} \right).
\end{equation}
Indeed, these three generators form a closed set of commutation
relations:
\begin{equation}\label{115}
 \left[J_{2}, K_{1}\right] = - iK_{3},  \qquad
 \left[J_{2}, K_{3}\right] = iK_{1},   \qquad
 \left[K_{1}, K_{3}\right] = iJ_{2}.
\end{equation}
This closed set is called the Lie algebra of the $Sp(2)$ group, isomorphic
to the Lorentz group applicable to two space  and one time dimensions.

Let us consider the Minkowskian space of $(x, y, z, t)$.  It is possible to
write three four-by-four matrices satisfying the Lie algebra of Eq.(\ref{115}).
The three four-by-four matrices satisfying this set
of commutation relations are:
\begin{equation}\label{119}
J_{2} = \pmatrix{0 & 0 & i & 0 \cr 0 & 0 & 0 & 0 \cr i & 0 & 0 & 0 \cr
    0 & 0 & 0 & 0 },\quad
K_{1} = \pmatrix{0 & 0 & 0 & i \cr 0 & 0 & 0 & 0 \cr 0 & 0 & 0 & 0 \cr
    i & 0 & 0 & 0 },  \quad
 K_{3} = \pmatrix{0 & 0 & 0 & 0 \cr 0 & 0 & 0 & 0 \cr 0 & 0 & 0 & i \cr
    0 & 0 & i & 0 } .
\end{equation}
However, these matrices have null second rows and null second columns.
Thus, they can generate Lorentz transformations applicable only to the
three-dimensional space of $(x,z,t)$, while the $y$ variable remains invariant.

\section{Two-oscillator System}\label{2osc}
In order to generate Lorentz transformations applicable to the full
Minkowskian space, along with $J_{2},K_{1}$, and $K_{3}$ we need two
more Heisenberg commutation relations.
Indeed, Paul A. M. Dirac started this program in 1963~\cite{dir63}.
It is possible to write the two uncertainty relations using two harmonic
oscillators as
\begin{equation}\label{121}
\left[a_{i}, a^{\dag}_{j}\right] = \delta_{ij} .
\end{equation}
with
\begin{equation}\label{123}
  a_{i} = \frac{1}{\sqrt{2}}\left(x_{i} + ip_{i} \right), \qquad
  a^{\dag}_{i} = \frac{1}{\sqrt{2}}\left(x_{i} - ip_{i} \right),
\end{equation}
and
\begin{equation}\label{125}
  x_{i} = \frac{1}{\sqrt{2}}\left(a_{i} + a^{\dag}_{i} \right), \qquad
  p_{i} = \frac{i}{\sqrt{2}}\left(a^{\dag}_{i} - a_{i} \right),
\end{equation}
 where $i$ and $j$ could be 1 or 2.

More recently in 1986, this two-oscillator system was considered
by Yurke {\it et al.}~\cite{yurke86} in their study of two-mode
interferometers.  They considered first
\begin{equation}\label{301}
 Q_{3} = \frac{i}{2}\left(a_{1}^{\dag}a_{2}^{\dag} - a_{1}a_{2}\right),
\end{equation}
which leads to the generation of the two-mode coherent state or the
squeezed state~\cite{yuen76}.

Yurke {\it et al.} then considered possible interferometers requiring
the following two additional operators.
\begin{equation}\label{303}
S_{3} = {1\over 2}\left(a^{\dag}_{1}a_{1} + a_{2}a^{\dag}_{2}\right) ,
\qquad
K_{3} = {1\over 2}\left(a^{\dag}_{1}a^{\dag}_{2} + a_{1}a_{2}\right) .
\end{equation}
The three Hermitian operators from Eq.(\ref{301}) and Eq.(\ref{303})
satisfy the commutation relations
\begin{equation} \label{305}
\left[K_{3}, Q_{3}\right] = -iS_{3}, \qquad
\left[Q_{3}, S_{3}\right] = iK_{3}, \qquad
\left[S_{3}, K_{3}\right] = iQ_{3} .
\end{equation}
These relations are like those given in Eq.(\ref{115}) for the Lorentz
group applicable to two space-like and one time-like dimensions.

In addition, in the same paper~\cite{yurke86}, Yurke {\it et al.}
discussed the possibility of constructing interferometers exhibiting
the symmetry generated by
\begin{equation}\label{307}
 L_{1} = {1\over 2}\left(a^{\dag}_{1}a_{2} + a^{\dag}_{2}a_{1}\right) , \quad
 L_{2} = {1\over 2i}\left(a^{\dag}_{1}a_{2} - a^{\dag}_{2}a_{1}\right), \quad
 L_{3} = {1\over 2}\left(a^{\dag}_{1}a_{1} - a^{\dag}_{2}a_{2} \right).
\end{equation}
These generators satisfy the closed set of commutation relations
\begin{equation}\label{309}
\left[L_{i}, L_{j}\right] = i\epsilon_{ijk} L_{k} ,
\end{equation}
and therefore define a Lie algebra which is the same as that for $SU(2)$
or the three-dimensional rotation group.

We are then led to ask whether it is possible to construct a closed set
of commutation relations with the six Hermitian operators from
Eqs.(\ref{301},\ref{303},\ref{307}). It is not possible.  We have to
add four additional operators, namely
\begin{eqnarray}\label{311}
&{}& K_{1} = -{1\over 4}\left(a^{\dag}_{1}a^{\dag}_{1} + a_{1}a_{1} -
  a^{\dag}_{2}a^{\dag}_{2} - a_{2}a_{2}\right) ,   \quad
K_{2} = +{i\over 4}\left(a^{\dag}_{1}a^{\dag}_{1} - a_{1}a_{1} +
  a^{\dag}_{2}a^{\dag}_{2} - a_{2}a_{2}\right) ,   \nonumber \\[1ex]
&{}& Q_{1} = -{i\over 4}\left(a^{\dag}_{1}a^{\dag}_{1} - a_{1}a_{1} -
  a^{\dag}_{2}a^{\dag}_{2} + a_{2}a_{2} \right) ,   \quad
Q_{2} = -{1\over 4}\left(a^{\dag}_{1}a^{\dag}_{1} + a_{1}a_{1} +
   a^{\dag}_{2}a^{\dag}_{2} + a_{2}a_{2} \right) .
\end{eqnarray}

There are now ten operators from Eqs.(\ref{301},\ref{303},\ref{307},\ref{311}).
Indeed, these ten operators satisfy the following closed set of commutation
relations.
\begin{eqnarray}\label{313}
&{}& [L_{i}, L_{j}] = i\epsilon _{ijk} L_{k} ,\quad
[L_{i}, K_{j}] = i\epsilon_{ijk} K_{k} ,  \quad
[L_{i}, Q_{j}] = i\epsilon_{ijk} Q_{k} , \nonumber\\[1ex]
 &{}& [K_{i}, K_{j}] = [Q_{i}, Q_{j}] = -i\epsilon _{ijk} L_{k} , \quad
             [K_{i}, Q_{j}] = -i\delta_{ij} S_{3},   \nonumber\\[1ex]
&{}&[L_{i}, S_{3}] = 0, \quad [K_{i}, S_{3}] =  -iQ_{i},\quad
[Q_{i}, S_{3}] = iK_{i} .
\end{eqnarray}
As Dirac noted in 1963~\cite{dir63}, this set is the same as the Lie
algebra for the $SO(3,2)$ de Sitter group, with ten generators. This
is the Lorentz group applicable to the three-dimensional space with
two time variables.  This group plays a very important role in
space-time symmetries.

In the same paper, Dirac pointed out that this set of commutation
relations serves as the Lie algebra for the four-dimensional
symplectic group commonly called $Sp(4)$, applicable to the systems
of two one-dimensional particles, each with a two-dimensional phase
space.

For a dynamical system consisting of two pairs of canonical variables
$x_{1}, p_{1}$ and $x_{2}, p_{2}$, we can use the four-dimensional
space with the coordinate variables defined as~\cite{hkn95jmp}
\begin{equation}\label{317}
\left(x_{1}, p_{1}, x_{2}, p_{2} \right).
\end{equation}
Then the four-by-four transformation matrix $M$ applicable to this
four-component vector is canonical if~\cite{abra78,goldstein80}
\begin{equation}\label{319}
M J \tilde{M} = J ,
\end{equation}
where $\tilde{M}$ is the transpose of the $M$ matrix,
with
\begin{equation}\label{321}
J = \pmatrix{0 & 1 & 0 & 0 \cr -1 & 0 & 0 & 0 \cr
             0 & 0 & 0 & 1 \cr 0 & 0 & -1 & 0 } .
\end{equation}
According to this form of the $J$ matrix, the area of the phase space for
the $x_{1}$ and $p_{1}$ variables remains invariant, and the story is the
same for the phase space of $x_{2}$ and $p_{2}.$

We can then write the generators of the $Sp(4)$ group as
\begin{eqnarray}\label{322}
&{}& L_{1} = -\frac{1}{2}\pmatrix{0 & I\cr I & 0 }\sigma_{2}  , \quad
L_{2} = \frac{i}{2} \pmatrix{0 & -I \cr I & 0} I , \quad
L_{3} = \frac{1}{2}\pmatrix{-I & 0 \cr 0 & I}\sigma_{2} , \nonumber\\[2ex]
&{}& S_{3} = \frac{1}{2}\pmatrix{I   & 0\cr 0 & I} \sigma_{2},
\end{eqnarray}
\noindent and
\begin{eqnarray}\label{323}
&{}&K_{1} = \frac{i}{2}\pmatrix{I  & 0 \cr 0 & -I } \sigma_{1}, \quad
K_{2} = \frac{i}{2} \pmatrix{I & 0 \cr 0 & I } \sigma_{3}, \quad
K_{3} = -\frac{i}{2}\pmatrix{0 & I \cr I & 0 }  \sigma_{1}, \nonumber\\[2ex]
&{}& Q_{1} = -\frac{i}{2}\pmatrix{I & 0 \cr 0 & -I }\sigma_{3}, \quad
Q_{2} = \frac{i}{2}\pmatrix{I & 0 \cr 0 & I }\sigma_{1} , \quad
Q_{3} = \frac{i}{2}\pmatrix{0 &  I \cr I  & 0 } \sigma_{3} ,
\end{eqnarray}
where $I$ is the two-by-two identity matrix, while $\sigma_{1},
\sigma_{2}$, and $\sigma_{3}$ are the two-by-two Pauli matrices.  The four matrices
given in Eq.(\ref{322}) generate rotations, while those of Eq.(\ref{323}) lead to
squeezes in the four-dimensional phase space.

As for the difference in methods used in Secs.~\ref{sp2} and~\ref{2osc},
let us look at the ten four-by-four matrices given in Eqs.(\ref{322}) and (\ref{323}).
Among these ten matrices, six of them are diagonal.  They are
$S_{3}, L_{3}, K_{1}, K_{2} , Q_{1},$ and $Q_{2}$.
In the language of two harmonic oscillators, these generators do not
mix up the first and second oscillators.   There are six of them because
each operator has three generators for its own $Sp(2)$ symmetry.
Let us consider the three generators, $S_{3}, K_{2}$, and $Q_{2}$.
For each oscillator, the generators consist of
\begin{equation}
  \sigma_{2}, \quad  i\sigma_{1} \quad \mbox{and} \quad i\sigma_{3} .
\end{equation}
These separable generators thus constitute the Lie algebra of Sp(2) group
for the one-oscillator system, which we discussed in Sec.~\ref{sp2}.  Hence, the
one-oscillator system constitutes a subgroup of the two-oscillator system.

The off-diagonal matrix $L_{2}$ couples the first and second oscillators
without changing the overall volume of the four-dimensional phase space.
However, in order to construct the closed set of commutation relations,
we need the three additional generators: $L_{1} , K_{3},$ and $Q_{3}.$
The commutation relations given in Eq.(\ref{313}) are clearly
consequences of Heisenberg's uncertainty relations.

\section{Contraction of SO(3,2) to ISO(3,1)}\label{contraction}

Let us next go back to the $SO(3,2)$ contents of this two-oscillator
system~\cite{dir63}.  There are three space-like coordinates $(x, y, z)$
and two time-like coordinates $s$ and $t$.  It is thus possible to
construct the five-dimensional space of $(x, y, z, t, s)$, and to consider
four-dimensional Minkowskian subspaces consisting of $(x, y, z, t)$
and $(x, y, z, s)$.

As for the $s$ variable, we can make it longer or shorter, according
to procedure of group contractions introduced first by In{\"o}n{\"u} and
Wigner~\cite{inonu53}.  In this five-dimensional space, the boosts along the
$x$ direction with respect to the $t$ and $s$ variables are generated by
\begin{equation}\label{333}
A_{x} = \pmatrix{0 & 0 & 0 & i & 0 \cr 0 & 0 & 0 & 0 & 0 \cr
0 & 0 & 0 & 0 & 0 \cr i & 0 & 0 & 0 & 0 \cr 0 & 0  & 0 & 0 & 0 }, \qquad
B_{x} = \pmatrix{0 & 0 & 0 & 0 & i \cr 0 & 0 & 0 & 0 & 0 \cr
0 & 0 & 0 & 0 & 0 \cr 0 & 0 & 0 & 0 & 0 \cr i & 0  & 0 & 0 & 0 },
\end{equation}
respectively.  The boost generators along the $y$ and $z$ directions
take similar forms.

Let us then introduce the five-by-five contraction matrix~\cite{kiwi87jmp,kiwi90jmp}
\begin{equation}\label{335}
C(\epsilon) = \pmatrix{1 & 0 & 0 & 0 & 0 \cr 0 & 1 & 0 & 0 & 0 \cr
0 & 0 & 1 & 0 & 0 \cr 0 & 0 & 0 & 1 & 0 \cr 0 & 0  & 0 & 0 &  \epsilon } .
\end{equation}
This matrix leaves the first four columns and rows invariant, and the
four-dimensional Minkowskian sub-space of $(x, y, z, t)$ stays invariant.

As for the boost with respect to the $s$ variable, according to the procedure
spelled out in Ref.~\cite{kiwi87jmp,kiwi90jmp}, the contracted boost generator becomes
\begin{equation} \label{341}
  B^{c}_{x} = \lim_{\epsilon\rightarrow \infty}\frac{1}{\epsilon}~
  \left[C^{-1}(\epsilon)~B_{x}~C(\epsilon)\right] =
   \pmatrix{0 & 0 & 0 & 0 &  i \cr 0 & 0 & 0 & 0 & 0 \cr
0 & 0 & 0 & 0 & 0 \cr 0 & 0 & 0 & 0 & 0  \cr 0 & 0 & 0 & 0 & 0 } .
\end{equation}
Likewise, $B^{c}_{y}$ and $B^{c}_{z}$ become
\begin{equation}\label{343}
     B^{c}_{y} = \pmatrix{0 & 0 & 0 & 0 &  0 \cr 0 & 0 & 0 & 0 & i \cr
     0 & 0 & 0 & 0 & 0 \cr 0 & 0 & 0 & 0 & 0  \cr 0 & 0 & 0 & 0 & 0 },
     \qquad
     B^{c}_{z} = \pmatrix{0 & 0 & 0 & 0 &  0 \cr 0 & 0 & 0 & 0 & 0 \cr
     0 & 0 & 0 & 0 & i \cr 0 & 0 & 0 & 0 & 0  \cr 0 & 0 & 0 & 0 & 0} ,
\end{equation}
respectively.

\begin{figure}
\centerline{\includegraphics[scale=2.0]{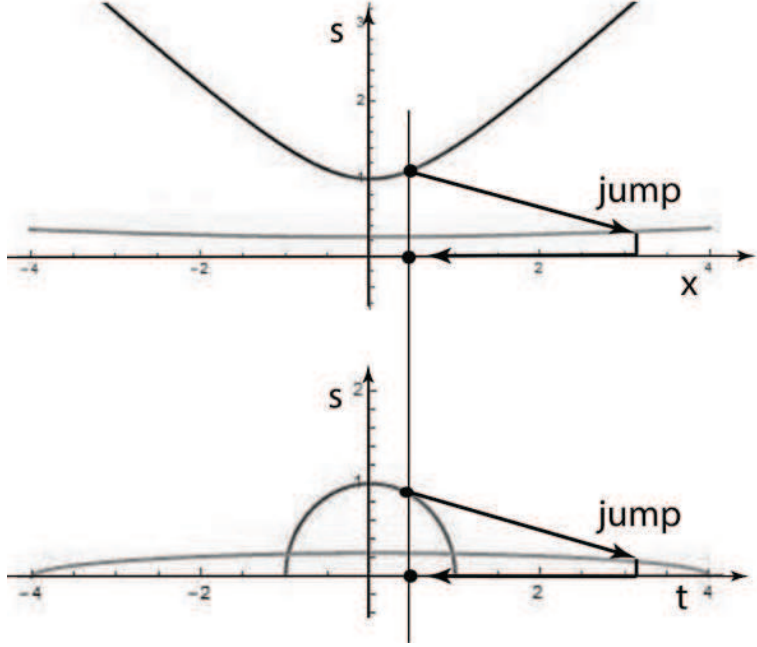}}
\caption{Contraction of the $SO(3,2)$ group to the Poincar\'e group.
The time-like $s$ coordinate is contracted with respect to the
space-like $x$ variable, and with respect to the time-like
variable $t$.}\label{contrac}
\end{figure}

As for the $t$ direction, the transformation applicable to the $s$ and $t$
variables is a rotation, generated by
\begin{equation}\label{345}
B_{t} = \pmatrix{0 & 0 & 0 & 0 & 0 \cr 0 & 0 & 0 & 0 & 0 \cr
0 & 0 & 0 & 0 & 0 \cr 0 & 0 & 0 & 0 & i \cr 0 & 0  & 0 & -i & 0 } .
\end{equation}
This matrix also becomes contracted to
\begin{equation}\label{347}
B^{c}_{t} = \pmatrix{0 & 0 & 0 & 0 & 0 \cr 0 & 0 & 0 & 0 & 0 \cr
0 & 0 & 0 & 0 & 0 \cr 0 & 0 & 0 & 0 & i \cr 0 & 0  & 0 & 0 & 0 } .
\end{equation}
These contraction procedures are illustrated in Fig.~\ref{contrac}.


These four contracted generators lead to the five-by-five transformation
matrix
\begin{equation}\label{349}
\exp\left\{-i\left(aB^{c}_{x}+ bB^{c}_{y} + cB^{c}_{z} + dB^{c}_{t}\right)\right\}
= \pmatrix{1 & 0 & 0 & 0 & a \cr 0 & 1 & 0 & 0 & b \cr
0 & 0 & 1 & 0 & c \cr 0 & 0 & 0 & 1 & d \cr 0 & 0  & 0 & 0 & 1 } ,
\end{equation}
performing translations:
\begin{equation}\label{351}
     \pmatrix{1 & 0 & 0 & 0 & a \cr 0 & 1 & 0 & 0 & b \cr
0 & 0 & 1 & 0 & c \cr 0 & 0 & 0 & 1 & d \cr 0 & 0  & 0 & 0 & 1 }
\pmatrix{x \cr y \cr z \cr t \cr 1 } =
\pmatrix{x + a \cr y + b \cr z + c \cr t + d \cr 1 } .
\end{equation}
This matrix leaves the first four rows and columns invariant.  They are for
the Lorentz transformation applicable to the Minkowskian space of $(x, y, z, t)$.

In this way, the boosts along the $s$ direction become contracted to the
translation.  This means the group $SO(3,2)$ derivable from the Heisenberg's
uncertainty relations becomes the inhomogeneous Lorentz group governing the
Poincar\'e symmetry for quantum mechanics and quantum field
theory in the Lorentz-covariant world~\cite{dir63,dir62}.

The group contraction has a long history in physics, starting from the 1953
paper by In{\"o}n{\"u} and Wigner~\cite{inonu53}.   It starts with a geometrical
concept.  Our earth is a sphere, but is is convenient to consider a flat surface
tangent to a given point on the spherical surface of the earth.  This approximation
is called the contraction of $SO(3)$ to $SE(2)$ or the two-dimensional Euclidean
group with one rotational and two translational degrees of freedom.

This mathematical method was extended to the contraction of the $SO(3,1)$ Lorentz
group to the three-dimensional Euclidean group.  More recently, Kim and Wigner
considered a cylindrical surface tangent to the sphere~\cite{kiwi87jmp,kiwi90jmp}
at its equatorial belt.  This cylinder has one rotational degree of freedom and
one up-down translational degree of freedom.  It was shown that the rotation and
translation correspond to the helicity and gauge degrees of freedom for massless
particles.

Since the Lorentz $SO(3,1)$ is isomorphic to the $SL(2,c)$ group of two-by-two
matrices, we can ask whether it is possible to perform the same contraction
procedure in the regime of two-by-two matrices.  It does not appear possible
to represent the $ISE(3)$ (inhomogeneous Euclidean group) with two-by-two
matrices.  Likewise, there seem to be difficulties in addressing the
question of contracting $SO(3,2)$ to $ISO(3,1)$ within the frame work of
the four-by-four matrices of $Sp(4)$.

\section{Concluding Remarks}\label{remarks}

Special relativity and quantum mechanics served as the major
theoretical basis for modern physics for one hundred years.
They coexisted in harmony: quantum mechanics augmented by Lorentz
covariance when needed.
Indeed, there have been attempts in the past to construct a
Lorentz-covariant quantum world by augmenting the Lorentz group
to the uncertainty relations~\cite{dir27,dir45,dir49,dir62,yuka53}.
There are recent papers on this subject~\cite{fkr71,lowe83,bars09}.
There are also papers on group contractions including contractions
of the $SO(3,2)$ group~\cite{gilmore74,bohm84,bohm85}.

It is about time for us to examine whether  both of these two great
theories can be synthesized.  The first step toward this process is
to find the common mathematical ground.  Before Newton, open orbits
(comets) and closed orbits (planets) were treated differently, but
Newton came up with one differential equation for both.  Before
Maxwell, electricity and magnetism were different branches of physics.
Maxwell's equations synthesized these two branches into one.  It is
shown in this paper that the group $ISO(3,1)$ can be derived from
the algebra of quantum mechanics.

It is gratifying to note that the Poincar\'e symmetry is derivable
within the system of Heisenberg's uncertainty relations.  The
procedure included two coupled oscillators resulting in the $SO(3,2)$
symmetry~\cite{dir63}, and the contraction of this $SO(3,2)$ to the
inhomogeneous Lorentz group $ISO(3,1)$.


\end{document}